\documentstyle[12pt]{article}
\begin{document}
\title{
\hfill\parbox{4cm} {\normalsize hep-th - 9906247}\\
Supergravity Solution for Three-String Junction 
in M-theory}
\author{P. Ramadevi
{\footnote{E-mail: rama@theory.tifr.res.in, 
ramadevi@phy.iitb.ernet.in}}\\
Physics Department,
Indian Institute of Technology Bombay,\\
Mumbai - 400 076, INDIA}
\date {}
\maketitle
\begin{abstract}
Three-String junctions are allowed configurations in 
II B string theory which preserve one-fourth supersymmetry.
We obtain the 11-dimensional supergravity solution for curved 
membranes corresponding to these three-string junctions.
\end{abstract} 

\def\vk{{\vec k}}
\def\ben{\begin{equation}}
\def\een{\end{equation}}
\def\bea{\begin{eqnarray}}
\def\eea{\end{eqnarray}}
\def\bean{\begin{eqnarray*}}
\def\eean{\end{eqnarray*}}
In the last few years, there has been lot of interest on 
three-string junctions in Type II B string theory \cite {schw}- \cite{hasii}.
It was orginally postulated by Schwarz that 
string theory can also admit multi-junction/multi-pronged strings
as fundamental objects \cite {schw}.
In Ref. \cite {gabe}, it has been shown
that the gauge field configurations in $F$-theory
on $K_3$ for  exceptional groups 
can be accounted only if multi-pronged strings besides
the usual strings (two-pronged) are 
included as solutions in string theory. The one-fourth BPS
nature of three-string junction was conjectured in
\cite {schw} and proven in \cite {mukh,asen}. 
These three-pronged strings/three-string junction
connecting three $D3$-branes 
gives understanding of ${1 \over 4}$-th 
BPS states of the $SU(3)$ super yang-mills 
theory on the coincident three- $D3$-branes\cite {berg,hasi}
and their $SU(N)$ generalisations presented in Ref. \cite {hasii}.

With enough evidence for such a junction configuration
to be present in string theory,
it is very essential that these junctions emerge as
exact solution of supergravity field equations.  
Till date, attempts in
10-dimensions have failed due to the singular
nature at the junction. The hope is that 
smoothening the junction would help in finding
the solution. Hence, we look at curved membranes in M-theory
corresponding to  three-string junctions in II B theory.
The supergravity solution for planar membranes corresponding to 
fundamental strings \cite{duff} gives us some idea in solving
the curved case. The solution for such curved membranes
representing three-string junction in IIB theory, after the 
forthcoming (tedious) calculations, are given in 
eqns.(\ref {zone}, \ref {ztwo}){\footnote {We came across a 
recent paper \cite {don} where a similar form appeared
for localised orthogonal intersecting
membranes.}}. It is important to stress that 
the mathematical limitations (analytic solution for a non-linear
differential equation) enables us to obtain only an implicit 
supergravity solution for three-string junction.

The two dimensional membrane corresponding to three-string 
junction is given by holomorphic curve \cite {krog}:
\ben
f(u,v) = \sqrt{(u-\lambda_1) (v - \lambda_2)} - \sqrt{\lambda_1 
\lambda_2} = 0~ \label {hol},
\een
where $u$ and $v$ are 
functions of complex coordinates $U$ and $V$ in $R^2 \times T^2$.

We begin by making an ansatz for the 11-dimensional metric $g_{MN}$ respecting 
$SO(6)$ symmetry and gauge fields $A_{MNP}$ ( $M,N,P= 0,1, \ldots 10$)
\begin{eqnarray}
ds^2 &=& -a_0(u,\bar u, v, \bar v,r) dt^2 + 2 g_{\mu \bar {\nu}}
(u,\bar u,v, \bar v,r) dx^{\mu} dx^{\bar {\nu}}\nonumber\\
~&~&+a_1(u, \bar u, v, \bar v,r) \sum_{i=5}^{10} (dx^m)^2~, \label {one}\\
A &=& A_{0 \mu \bar {\nu}}(u, \bar u, v, \bar v, r) dx^0 \wedge dx^{\mu} \wedge
\wedge dx^{\bar {\nu}} , \ \label {two}
\end{eqnarray}
where $r^2 = \sum_{m=5}^{10} (x^m)^2$ and $x^0=t$, 
$x^{\mu} = u,v$; $x^{\bar \nu}= \bar u, \bar v$. 

The vielbeins $e^{\hat a}_M$ and inverse vielbeins $E^M_{\hat a}$
for the metric ansatz in the upper triangular form:
\begin{eqnarray}
e^{\hat 0}_0&=&\sqrt{a_0}=(E^0_{\hat 0})^{-1}~\nonumber\\ 
e^{\hat m}_n&=& \delta_{mn} \sqrt{a_1}=(E^m_{\hat n})^{-1}~,\nonumber\\ 
e^{\hat v}_v&=& e^{\hat {\bar v}}_{\bar v}= \sqrt{g_{v \bar v}}
= (E^v_{\hat v})^{-1} =(E^{\bar v}_{\hat {\bar v}})^{-1}~,
\nonumber\\
e^{\hat u}_u &=& e^{\hat {\bar u}}_{\bar u}= \sqrt{g_{u \bar u} g_{v \bar v}
-g_{u \bar v} g_{v \bar u} \over g_{v \bar v}} \nonumber\\~
&=&(E^u_{\hat u})^{-1} = (E^{\bar u}_{\hat {\bar u}})^{-1}~,
\nonumber\\
e^{\hat v}_u &=& {g_{\bar v u} \over \sqrt{g_{v \bar v}}} ~;
e^{\hat {\bar v}}_{\bar u} = {g_{\bar u v} \over \sqrt{g_{v \bar v}}}~,
\nonumber\\ 
E^v_{\hat u}&=& -{g_{u \bar v} \over \sqrt{g_{v \bar v} (g_{u \bar u}
g_{v \bar v}- g_{u \bar v} g_{v \bar u})}}~,\nonumber\\
E^{\bar v}_{\hat {\bar u}} &=& -{g_{v \bar u} \over \sqrt{g_{v \bar v} 
(g_{u \bar u} g_{v \bar v}- g_{u \bar v} g_{v \bar u})}}~,
\end{eqnarray}
where the indices with hat refers to tangent space index distinguishing
them from world volume indices. The  arbitrary functions
$a_i$'s and three-form components must be reduced to fewer number
of unknowns  by requiring that the field configuration (\ref {one}, \ref{two})
preserve one-fourth supersymmetry. In other words, there 
must exist Killing spinors satisfying
\ben
D_M \epsilon = 0 \label {thre}
\een
where $D_M$ is the supercovariant derivative appearing in 
the gravitino supersymmetry transformation,
\begin{eqnarray}
\delta \psi_M& = &D_M \epsilon, \\
D_M &=& \partial_M + {1 \over 4} \omega_M^{\hat A \hat B} \Gamma_{\hat A 
\hat B} \nonumber\\
~&~&- {1 \over 288} (\Gamma^{PQRS}_M + 8 \Gamma^{PQR} \delta^S_M)
F_{PQRS}
\end{eqnarray}
where $F_{MNPQ}= 4 \partial_{[M}A_{NPQ]}$. Here $\Gamma_{\hat A}$ 
are the $D=11$ Dirac matrices obeying
\ben
\{ \Gamma_{\hat A} , \Gamma_{\hat B} \} = 2 \eta_{\hat A \hat B},
\een
where $\eta_{\hat A \hat B}= {\rm diag}(-,+,+ \ldots +)$
and 
\ben
\Gamma_{\hat A \hat B \ldots \hat C} = \Gamma_{[\hat A} \Gamma_{\hat B}
\ldots \Gamma_{\hat C]}
\een
and $\Gamma$'s with world volume index can be converted to
tangent space index using vielbeins. 

We will split the 11-dimensional $\Gamma$ matrix respecting $S0(6)$ 
symmetry in the following way: 
\begin{equation}
\Gamma_{\hat A} = (i\gamma_{\hat {\alpha}} \otimes \Gamma_7, 
1 \otimes \Sigma_{\hat a})
\end{equation}
where $\gamma_{\hat {\alpha}}$ and $\Sigma_{\hat a}$ are the 
$D=5$ and $D=6$ Dirac matrices and 
\ben
\Gamma_7 = \Sigma_{\hat 5} \Sigma_{\hat 6} \ldots \Sigma_{\hat {10}}~.
\een
satisfying the following properties:
\begin{equation}
\gamma_{\hat 0} \gamma_{{\hat u} {\hat {\bar u}}}
\gamma_{{\hat v} {\hat {\bar v}}} = i~;~
\Gamma_7^2 = -1
\end{equation}

To simplify the computations, we shall first impose the following
constraints on ${\epsilon}$ 
\begin{equation}
\gamma_u \epsilon  = 0 ~;~ 
\gamma_v \epsilon = 0 ~;~
\Gamma_7 \epsilon = -i \epsilon~, \label {four}
\end{equation}
which in the 11-dimensions gives one-fourth BPS  
nature of the curved membranes corresponding to three-string junction-viz.,
\begin{eqnarray}
\Gamma_1 \Gamma_2 \Gamma_5 \ldots \Gamma_{10} \epsilon = \epsilon~,\\
\Gamma_3 \Gamma_4 \Gamma_5 \ldots \Gamma_{10} \epsilon = \epsilon~.
\end{eqnarray} 
With the above constraints, the spin connection on the spinor field can be 
simplified to
\ben
(\omega_M)_{\hat A \hat B} \Gamma^{\hat A \hat B} \epsilon=
E^N_{\hat A} E^R_{\hat B} \partial_R g_{NM} \Gamma^{\hat A \hat B} 
\epsilon
\een
In our background (\ref {one}, \ref{two}), we shall now examine 
eqn.(\ref {thre}).
After incorporating one-fourth BPS condition (\ref {four}), we get  
\begin{eqnarray}
D_0 \epsilon= \partial_0 \epsilon + 
[{1 \over 4} \gamma_0 \gamma^u (\partial_u
\ln a_0) + {1 \over 4} \gamma_0 \gamma^v (\partial_v \ln a_0)~~~ \nonumber\\
~~- {1 \over 4} \gamma_0 \Sigma^m (\partial_m \ln a_0) 
- {i \over 6} \sqrt{a_0^{-1}} \gamma_0 \Sigma^m H\nonumber\\ 
~~+{i \over 6} \sqrt{a_0^{-1}} \gamma_0 I \gamma^u 
+{i \over 6} \sqrt{a_0^{-1}} \gamma_0 J \gamma^v] \epsilon
 = 0~~~\label {six}\\ 
D_m \epsilon = \partial_m \epsilon 
+[ {1 \over 8} (\Sigma_m \Sigma^n - \Sigma^n \Sigma_m)  
(\partial_n \ln a_1)~~~~~~~~~ \nonumber\\
~~+ {1 \over 4} \Sigma_m  \gamma^u (\partial_u \ln a_1)
+ {1 \over 4} \Sigma_m  \gamma^u (\partial_u \ln a_1)\nonumber\\ 
~~-{i \over 24} \sqrt{a_0^{-1}} ( 
  \Sigma_m \Sigma^n - \Sigma^n \Sigma_m ) H
 -  {i \over 12} \sqrt{a_0^{-1}} \Sigma_m \gamma^u I \nonumber\\
~~- {i \over 12} \sqrt{a_0^{-1}} \Sigma_m \gamma^v J 
~~+{i \over 6} 
\sqrt{a_0^{-1}} H] \epsilon = 0~,\label{seve}\\
D_u \epsilon= \partial_u \epsilon 
+[{1 \over 4} \left( g^{u \bar u} \partial_u g_{u \bar u}
+ g^{v \bar v} \partial_v g_{\bar v u} \right. ~~~~~~~~~~~~~~~~\nonumber\\
~~~\left. + g^{\bar u v}
\partial_v g_{u \bar u} + g^{u \bar v} \partial_u g_{\bar v u} \right)
+{i \sqrt{a_0^{-1}}\over 12}  I] \epsilon=0~,  \label {unbaru}\\
D_v \epsilon= \partial_v \epsilon  
+[{1 \over 4} \left( g^{v \bar v} \partial_v g_{v \bar v}
+ g^{u \bar v} \partial_u g_{\bar v v}\right.~~~~~~~~~~~~~ \nonumber\\
~~~\left. + g^{\bar u u}
\partial_u g_{v \bar u} + g^{v \bar u} \partial_v g_{\bar u v} \right)
+{i \sqrt{a_0^{-1}}\over 12}  J] \epsilon=0~, \label {unbarv}\\
D_{\bar u} \epsilon=\partial_{\bar u} \epsilon
-[{1 \over 4} \left(g^{u \bar u} \partial_{\bar u} g_{u \bar u}
+ g^{u \bar v} \partial_{\bar v} g_{u \bar u} + g^{v \bar v}
\partial_{\bar v} g_{v \bar u} \right. \nonumber\\
~~\left. + g^{v \bar u} \partial_{\bar u}
g_{v \bar u} \right)
- {1 \over 4} \Sigma^m \gamma^v \partial_m g_{v \bar u} 
- {1 \over 4} \Sigma^m \gamma^u \partial_m g_{u \bar u}\nonumber\\ 
~~+{1 \over 4} \gamma^{u v} \left( \partial_v g_{u \bar u}
- \partial_u g_{v \bar u} \right) 
 - \Sigma^m \gamma^v {i \over 12 } \sqrt{a_0^{-1}} 
( g^{\bar v u} P  \nonumber\\
~~- g^{u \bar u} Q)
+  \Sigma^m \gamma^u {i \over 12 } \sqrt{a_0^{-1}} 
( g^{\bar v v} P - g^{v \bar u} Q) \nonumber\\ 
~~ + {i \over 6} \sqrt{a_0^{-1}}\left(-\Sigma^m 
\gamma^u  \partial_m A_{0 u \bar u} 
+ \Sigma^m \gamma^v  \partial_m A_{o \bar u v}\right.\nonumber\\ 
~~~\left. + \gamma^{uv}\{
\partial_u A_{0 \bar u v } + \partial_v A_{0 u \bar u} \}\right) 
+{i \over 4} \sqrt{a_0^{-1}} K ] \epsilon=0~, \label{baru}\\
D_{\bar v} \epsilon = \partial_{\bar v} \epsilon
-[{1 \over 4} \left(g^{v \bar v} \partial_{\bar v} g_{v \bar v}
+ g^{u \bar v} \partial_{\bar v} g_{u \bar v} + g^{u \bar u}
\partial_{\bar u} g_{u \bar v}\right. \nonumber\\
~~~\left.+ g^{v \bar u} \partial_{\bar u}
g_{v \bar v} \right)
- {1 \over 4} \Sigma^m \gamma^v \partial_m g_{v \bar v} 
- {1 \over 4} \Sigma^m \gamma^u \partial_m g_{u \bar v}\nonumber\\ 
~~~ -{1 \over 4} \gamma^{u v} \left( \partial_u g_{v \bar v}
- \partial_v g_{u \bar v} \right)
 - \Sigma^m \gamma^u {i \over 12 } \sqrt{a_0^{-1}} 
( g^{\bar u v} R \nonumber\\
~- g^{v \bar v} S)
+  \Sigma^m \gamma^v {i \over 12 } \sqrt{a_0^{-1}} 
( g^{\bar u u} R - g^{u \bar v} S) \nonumber\\ 
~~~ +  {i \over 6} \sqrt{a_0^{-1}}\left( \Sigma^m \gamma^u 
\partial_m A_{0 \bar v u} 
- \Sigma^m \gamma^v  \partial_m A_{0 v \bar v} \right.\nonumber\\
~~~ \left.- \gamma^{uv}\{
\partial_u A_{0 v \bar v} + \partial_v A_{0 \bar v u} \} \right)
+{i \sqrt{a_0^{-1}} \over 4}L 
] \epsilon=0~, \label {barv}
\end{eqnarray}
where 
\begin{eqnarray}
H &=& g^{u \bar u} \partial_m A_{u \bar u 0} + g^{u \bar v}
\partial_m A_{u \bar v 0} + g^{v \bar u} \partial_m A_{v \bar u 0}\nonumber\\
~&~&+ g^{v \bar v} \partial_m A_{v \bar v 0}\\
I &=& g^{v \bar v} \partial_u A_{v \bar v 0} - g^{v \bar u} 
\partial_u A_{\bar u v 0} + g^{v \bar v} \partial_v A_{\bar v u 0}\nonumber\\
~&~&-g^{\bar u v} \partial_v A_{u \bar u 0}\\
J &=& g^{u \bar u} \partial_u A_{\bar u v 0} - g^{u \bar v} 
\partial_u A_{v \bar v 0} + g^{u \bar u} \partial_v A_{u \bar u 0}\nonumber\\
~&~&-g^{\bar v u} \partial_v A_{\bar v u 0}\\
P&=& \left(g_{u \bar u} \partial_m A_{v \bar v 0}
- g_{\bar u v} \partial_m A_{u \bar v 0} \right)\\
Q&=& \left(g_{u \bar u} \partial_m A_{\bar u v 0}
+ g_{\bar u v} \partial_m A_{u \bar u 0} \right)\\
K&=&\left(g^{v \bar v} \partial_{\bar v} A_{0 \bar u v} - g^{\bar v u} 
\partial_{\bar v} A_{0 u \bar u} +g^{v \bar v} 
\partial_{\bar u} A_{v \bar v 0 }\right. \nonumber\\
~&~&\left.- g^{\bar v u} \partial_{\bar u} A_{\bar v u 0 } \right)\\ 
R&=& \left(g_{v \bar v} \partial_m A_{u \bar u 0}
- g_{\bar v u} \partial_m A_{v \bar u 0} \right)\\
S&=& \left( g_{v \bar v} \partial_m A_{\bar v u 0} + 
g_{\bar v u} \partial_m A_{v \bar v 0} \right)\\ 
L&=&\left (g^{u \bar u} 
\partial_{\bar u} A_{u 0 \bar v} - g^{\bar u v} 
\partial_{\bar u} A_{0 v \bar v} 
+g^{u \bar u} \partial_{\bar v} A_{u \bar u 0 }\right. \nonumber\\
~&~&\left. - g^{\bar u v} 
\partial_{\bar v} A_{\bar u v 0 } \right) 
\end{eqnarray}
From eqn.(\ref{six}), equating the respective $\Gamma$ terms 
we get,
\begin{eqnarray}
\partial_0 \epsilon&=&0~, \label {eiga} \\ 
\gamma_0 \gamma^u ({1 \over 4} \partial_u {\rm ln} a_0 + {i \over 6}
\sqrt{a_0^{-1}} I ) &=& 0 ~, \label {eigh}\\
\gamma_0 \gamma^v ({1 \over 4} \partial_v {\rm ln} a_0 + {i \over 6}
\sqrt{a_0^{-1}} J ) &=& 0~, \label {eigha}\\
\gamma_0 \Sigma^m ( - {1 \over 4} \partial_m {\rm ln} a_0 -  {i \over 6} 
\sqrt{a_0^{-1}} H ) &=& 0~. \label {eighb}
\end{eqnarray}
Similarly equating the respective $\Gamma$ terms 
in (\ref{seve}) we get:
\begin{eqnarray}
\partial_n \epsilon&=& 
- {i \over 6} \sqrt{a_0^{-1}} H \epsilon  ~,\label{aone}\\ 
{1 \over 4} \partial_u {\rm ln} a_1 &=& {i \over 12} \sqrt{a_0^{-1}}
I~, \label {nine}\\ 
{1 \over 4} \partial_v {\rm ln} a_1 &=& {i \over 12} \sqrt{a_0^{-1}}
J~, \label {ninea}\\ 
{1 \over 8} \partial_n {\rm ln} a_1 &=& {i \over 24 } \sqrt{a_0^{-1}} H
~. \label {nineb} 
\end{eqnarray}
Comparing the above equations with 
eqns. (\ref{eigh}, \ref {eigha}, \ref {eighb}), we deduce  
\begin{eqnarray}
\partial_n \epsilon&=&{1 \over 4} (\partial_n {\rm ln} a_0) \epsilon ~, 
\label{atwo}\\ 
\partial_m {\rm ln}\sqrt{a_0^{-1}}&=& \partial_m {\rm ln} a_1 \label {ten}\\
\partial_{\mu} {\rm ln}\sqrt{a_0^{-1}}&=& \partial_{\mu} {\rm ln} a_1~
~, \label {elev}
\end{eqnarray} 
suggesting a relation
\begin{equation}
a_1 = \sqrt{a_0^{-1}}~. \label {rzer}
\end{equation}
Clearly, we have not used the actual form of $g_{\mu \bar {\nu}}$
and $A_{0 \mu \bar {\nu}}$ in deducing the relation between $a_0$ and $a_1$.
We will see that the equations obtained by 
comparing $\Gamma$ terms in (\ref {baru}, 
\ref {barv}) will help us to determine three-form
components and $g_{\mu \bar {\nu}}$.
The set of equations we get from
equating the respective $\gamma$ terms in (\ref {baru}) are
\begin{eqnarray}
\partial_{\bar u} \epsilon 
-[{1 \over 4} \left(g^{u \bar u} \partial_{\bar u} g_{u \bar u}
+ g^{u \bar v} \partial_{\bar v} g_{u \bar u} + g^{v \bar v}
\partial_{\bar v} g_{v \bar u}\right.\nonumber\\
~~\left. + g^{v \bar u} \partial_{\bar u}
g_{v \bar u} \right)
+{i \over 4} \sqrt{a_0^{-1}} K ] \epsilon=0 ~,
 \label {kzer}\\ 
\Sigma^m \gamma^v\{ - {1 \over 4} \partial_m g_{v \bar u}  -  {i \over 12 }
\sqrt{a_0^{-1}} (g^{\bar v u} P - g^{u \bar u} Q) \nonumber\\ 
~~+  {i \over 6} \sqrt{a_0^{-1}} \partial_m A_{0 \bar u v}\}=0~,\label{kone}\\ 
\Sigma^m \gamma^u\{ - {1 \over 4} \partial_m g_{u \bar u} 
+ {i \over 12} \sqrt{a_0^{-1}}(g^{v \bar v} P - g^{\bar u v} Q)\nonumber\\ 
-{i \over 6} \sqrt{a_0^{-1}} \partial_m A_{0 u \bar u} \}=0~, \label {ktwo}\\ 
\gamma^{u v} \{ {1 \over 4} (\partial_v g_{u \bar u} - 
\partial_u g_{v \bar u}) 
+{i \over 6} \sqrt{a_0^{-1}} (\partial_u A_{0 \bar u v}\nonumber\\
 + \partial_v A_{0 u \bar u}) \} =0 ~.\label {kthre}  
\end{eqnarray}
Similarly, we get the following set from eqn.(\ref {barv}):
\begin{eqnarray}
\partial_{\bar v} \epsilon
-[{1 \over 4} \left(g^{v \bar v} \partial_{\bar v} g_{v \bar v}
+ g^{u \bar v} \partial_{\bar v} g_{u \bar v} + g^{u \bar u}
\partial_{\bar u} g_{u \bar v}\right. \nonumber\\
\left. + g^{v \bar u} \partial_{\bar u}
g_{v \bar v} \right)
+{i \over 4} \sqrt{a_0^{-1}} L ] \epsilon=0 ~, \label {lzer}\\
\Sigma^m \gamma^u\{ - {1 \over 4} \partial_m g_{u \bar v}  -  {i \over 12 }
\sqrt{a_0^{-1}} (g^{\bar u v} R - g^{v \bar v} S)\nonumber\\ 
+  {i \over 6} \sqrt{a_0^{-1}} \partial_m A_{0 \bar v u}\}=0~,\label {lone}\\ 
\Sigma^m \gamma^v\{ - {1 \over 4} \partial_m g_{v \bar v} 
+ {i \over 12} \sqrt{a_0^{-1}}(g^{u \bar u} R - g^{\bar v u} S)\nonumber\\ 
-{i \over 6} \sqrt{a_0^{-1}} \partial_m A_{0 v \bar v} \}=0~, \label {ltwo}\\ 
\gamma^{u v} \{ -{1 \over 4} (\partial_u g_{v \bar v} - 
\partial_v g_{u \bar v}) 
-{i \over 6} \sqrt{a_0^{-1}} (\partial_u A_{0 v \bar v} \nonumber\\+ 
\partial_v A_{0 \bar v u }) \} =0 ~.\label {lthre}  
\end{eqnarray}
There are at least {\it three} solutions solving 
eqns. (\ref {eiga}- \ref{lthre}).

1) {\it For a sub-class of membranes satisfying
${\cal N} = \vert \partial_u f \vert ^2 + \vert\partial_v f \vert^2=
{\rm const}$}: 
\begin{eqnarray}
ds^2= H^{-2 \over 3}(r,\vert f\vert)\left(-dt^2 + 2 \vert du \vert^2
+2 \vert dv \vert^2 \right.~~~~~~~\nonumber\\ 
~~\left. - {2 \over {\cal N}}\vert df \vert^2 \right) 
+ H^{1 \over 3}(r, \vert f \vert) \left({2 \over {\cal N}} 
\vert df \vert^2 + 
\sum_{i=5}^{10} dx_m^2 \right)\\
A= {1 \over 2 {\cal N}} i H^{-1 \over 3} (r, \vert f \vert) 
\{-dt \wedge *\left(
df \wedge \bar d \bar f \right) \}~~~~~~~~~\\
\epsilon=\epsilon_0 H^{-1 \over 6}(r, \vert f \vert)~~~~~~~~~~~~~~
~~~~~~~~~~~~~~~~~~~~~~~~
\end{eqnarray}
where the Hodge star operation $*$ is 
done on the $R^2 \times T^2$ space.  
The three-form potential in component form for the above metric,
in the convention $\epsilon^{ u \bar u v \bar v}= +1$, simplifies to:
\begin{eqnarray}
A_{[0 u \bar u]} &=& i {1 \over {\cal N}} H^{-1} (r, \vert f \vert)  
\vert \partial_v f \vert^2~,~ \nonumber\\ 
A_{[0 u \bar v]} &=& - i {1 \over {\cal N}} H^{-1} (r, \vert f \vert)  
\partial_u f \partial_{\bar v} \bar f ~,\nonumber\\ 
A_{[0 v \bar v]}&=&{1 \over {\cal N}}  H^{-1} (r, \vert f \vert)  
\vert \partial_u f \vert^2~,~\nonumber\\ 
A_{[0 v \bar u]}& =& - i {1 \over {\cal N}} H^{-1} (r, \vert f \vert)  
\partial_v f \partial_{\bar u} \bar f  
\end{eqnarray}
This restricted class includes holomorphic curves $f=pu+qv$
representing planar membranes corresponding to
$(p,q)$ strings in IIB theory which preserve half supersymmetry. 
The membrane solution for $f=u$ and  $f=v$ agrees with 
the results in Ref.\cite {duff}.

2) {\it Intersecting $M2  \perp M2 $ branes at a point \cite {tsey,gaun}}:
\begin{eqnarray}
ds^2&=& H_1^{1 \over 3} (r) H_2^{1 \over 3} (r) \{ - H_1^{-1}(r) 
H_2^{-1} (r) dt^2 \nonumber\\
~&~&+ 2 H_1^{-1}(r) |du|^2
+ 2H_2^{-1}(r)|dv|^2
+\sum_{i=5}^{10} (dx_m)^2\}\\
A&=& i H_1^{-1} dt \wedge du \wedge d\bar u + i H_2^{-1} dt
\wedge dv \wedge d \bar v\\
\epsilon&=&\epsilon_0 H_1^{-1 \over 6} H_2^{-1 \over 6} \label {deloc}
\end{eqnarray}
where $\epsilon_0$ is a constant and $H_1(r)$, $H_2(r)$ are
harmonic functions dependent only on the transverse coordinates
common to both the branes.
These solutions are meaningful only if
$U,V$ are  compact coordinates  
with the charges being smeared over the branes.  

(a) For a coordinate transformation $u= x_1 + ix_4 ~;~v= x_2 + ix_3$, 
 the usual Kaluza-Klein reduction along $x_3$ on the above metric
and T-duality along $x_4$ gives the following ten-dimensional string metric:
\begin{eqnarray}
ds_{10}^2&=& -H_1^{-1 \over 2} (r) H_2^{-1 }(r)  dt^2 +
H_1^{-1 \over  2}(r)  dx_1^2 \nonumber\\ 
~&~&+ H_1^{1 \over 2}(r)  H_2^{-1}(r)  dx_2^2 + H_1^{1 \over 2}(r) 
 \sum_{m=4}^{10}( dx_m)^2~.
\end{eqnarray}
This represents delocalised solution for {\it orthogonal intersection of 
fundamental string along $x_2$ and  D string along $x_1$}. 

(b) For another coordinate transformation 
$u= (z_1 \tau_2 - \tau_1 z_2)~,~ 
v = z_2 ,$ 
where $z_1 = x_1 + ix_4$, $z_2 = x_2 + ix_3$, we
obtain the following metric after dimensional
reduction along $x_3$ and $T$-duality along
$x_4$:
\begin{eqnarray}
ds_{10}^2& =& B(r) \left( - H_1^{-1 \over 2}(r) H_2^{-1}(r) dt^2 + 
 H_1^{-1 \over 2}(r) \right. \nonumber\\
~&~&\left. (dx_1 \tau_2 - dx_2 \tau_1) ^2 + 
H_2^{-1}(r) H_1^{1 \over 2} (r) dx_2^2 \right.\nonumber\\
~&~&\left. + H_1^{1 \over 2}(r) \tau_2^{-2} dx_4^2
+ H_1^{1 \over 2} (r) 
\sum_{m=5}^{10} (dx_m)^2 \right)~,
\end{eqnarray}
where $B(r) = \sqrt{1+ H_1^{-1} (r) H_2 (r) \tau_1^2}$. This
solution is the simplest planar network of F-strings and D-strings
directed along $(0,1)$  and $(-\tau_2, \tau_1)$ in the $(x_1,x_2)$
plane. 

General planar network involving $[p_1,q_1], [p_2,q_2]$ strings can be 
similarly obtained using the coordinate transformation :
$u = p_1 (\tau_2 z_1 - \tau_1 z_2) - q_1 z_2~,~ v= q_2 z_2 - p_2
(\tau_2 z_1 - \tau_1 z_2)$. 

The equivalence of delocalised orthogonal intersecting strings
with general planar network of strings is expected because
the delocalised solution in $M$-theory (\ref {deloc})
has no information about the intersection point 
or the subspace containing the two $M2$-branes.
{\it Hence eqn. (\ref {deloc}) also represent
delocalised solutions for the membrane
corresponding to general planar network of strings.}

In order to distinguish the intersecting membranes from 
curved membranes corresponding to three-string junction,
we have to look for {\it fully or partially
localised solutions}.

3) {\it Arbitrary membranes including those corresponding 
to three-string junction}
 \begin{eqnarray}
ds^2&=&-H^{-2 \over 3} dt^2 +  2 H^{-2 \over 3} G_{\mu \bar {\nu}} 
(u, \bar u, v, \bar v, r) dx^{\mu}
dx^{\bar {\nu}}\nonumber\\
~&~&+ H^{1 \over 3} \sum_{m=5}^{10}(dx_m)^2 \label {zone}\\
A&=& i H^{-1} G_{\mu \bar {\nu}} dx_0 \wedge dx_{\mu} \wedge dx_{\bar {\nu}}
\label {ztwo} \\
\epsilon&=& \epsilon_0  H^{-1 \over 6}~, \label {zthr} 
\end{eqnarray} 
where $G_{\mu \bar {\nu}}$ is Kahler and the function $H$ is 
\begin{equation}
H=  G_{u \bar u} G_{v \bar v}- G_{u \bar v} G_{v \bar u}~. 
\end{equation}
The metric in terms of the Kahler potential $K$ is
\begin{equation}
G_{\mu \bar {\nu}} = \partial_{\mu} \partial_{\bar {\nu}} K~.
\end{equation}
The embedding of the membrane (\ref {hol}) in the metric can
be seen by performing the following 
holomorphic coordinate tranformation with unit Jacobian:
\begin{eqnarray}
(u,v) &\rightarrow& (\alpha, \beta)~, \\ 
{\rm where}~~ \alpha = \sqrt{(u- \lambda_1)
(v-\lambda_2)} &;&  
\beta= \alpha {\rm ln} {u - \lambda_1 \over v - \lambda_2}. \nonumber
\end{eqnarray}
The membrane surface in the new coordinate is
$f= \alpha - \sqrt{\lambda_1 \lambda_2}$.
This membrane is different from the intersecting membranes given by
$u = \lambda_1 ; v = \lambda_2$. We hope to see this difference
from partially  or fully localised supergravity solution.

We are now left with the task of determining the form of $K$
from the equations of motion for three-form gauge field 
in the presence of the curved membrane as the source:
\begin{eqnarray}
\partial_M(\sqrt{-g} F^{MUVW}) + {1 \over 1152}\left(  
\epsilon^{UVWMNOPQRST} \right. \nonumber\\
\left. F_{MNOP} F_{QRST} \right)
=J^{UVW}(x)\nonumber\\
={1 \over \sqrt{-g}} {\delta S_{membrane} \over \delta_{A_{UVW}(x)}}
\end{eqnarray}
where the membrane action is
\begin{eqnarray}
S_{membrane}&=&-T \int d^3 \xi \{ \sqrt{-{\rm det} h} \nonumber\\
~&~&- {1 \over 6} \epsilon^{a b c} \partial_a X^M \partial_b X^N
\partial_c X^P A_{MNP}\}~.
\end{eqnarray}
Here $h_{ab} = \partial_a X^M \partial_b X^N G_{MN}$ is the
induced metric on the membrane.

Choosing $\xi_0 =t, \xi_1= \beta, 
\xi_2 = \bar {\beta}$, we obtain the three-form current 
for the holomorphic membrane $f$ to be
\ben
J^{0 \beta \bar {\beta}} = {T \over \sqrt{-g}} \delta^2 (f) 
\delta^6 (x_m)~ \label {sour} . 
\een
Substituting the 3-form potential and metric (\ref {zone},
\ref {ztwo}) respecting one-fourth supersymmetry,
we get
\ben
\partial_{\mu} \partial_{\bar 
{\nu}} \left( 2H + \delta^{mn} \partial_m \partial_n K \right) =
j_{\mu \bar \nu} \label {nonl}
\een
where
\ben
j_{\mu \bar \nu} = \epsilon_{\mu \mu'} \epsilon_{\nu \nu'} J^{0 \mu' \nu'}
\sqrt{-{\rm det} g}~.
\een
The above non-linear equation cannot be solved analytically. Perturbative
approach over Minkowskian backgrounds gives integral representation 
for $K$ which is dependent on $j_{\mu \bar {\nu}}$ \cite {don}.
Since $j_{u \bar u} = \delta^2(u) \delta^6(r),~ j_{v \bar v} = 
\delta^2(v) \delta^6(r)$ for the intersecting membranes 
is  different from that of the curved membranes (\ref {sour}), 
the integrands of the integral representation for $K$
are distinct. 

In this approach \cite {don}, 
we expand the Kahler potential $K = \sum_n K^{(n)}$
and hence the metric $G_{m \bar n} = \sum_l G_{ m \bar n}^{(l)}$.
Minkowskian background implies that we 
take the zeroth-order $K^{(0)} = u \bar u + v \bar v$ so
that $G_{m \bar n}^{(0)} = \delta_{m \bar n}$. 
Further, comparing $n$-th order terms in eqn. (\ref {nonl}) 
gives a set of differential equations. Using this set, 
we get a formal integral representation for $K^{(n)}$ 
involving the lower order metric components. 
However, the goal to obtain explicit closed form 
expression for localised or partially localised solution
from the integral representation is still unsolved.  

It has been shown, 
for certain classes of orthogonal intersections
and holomorphic curves, 
that the perturbation theory
breaks down when there are more than three
transverse dimensions \cite {don}.  
This breakdown of the perturbation theory suggests that
no such fully localized solutions exist with asymptotically
flat boundary conditions. However, perturbation theory 
suggests that such solutions do exist when there are less than
three transverse dimensions. So, for sufficiently smeared 
versions of the sources, one expects 
that the solutions could be obtained numerically even if an 
exact analytic form cannot be found.

It is not clear at this stage whether perturbation
theory over other backgrounds like planar membrane 
background would help in finding a closed form expression. 
We hope to pursue this isssue in future. 

The fully localised or partially localised supergravity solutions
is also needed to understand the map of bulk parameters to 
boundary gauge theory parameters to prove AdS-CFT correspondence.
Such near-horizon or AdS metric for intersecting branes \cite {peet,youm}
and intersecting M5-branes \cite {faya} corresponding to NS5-D4 branes in 
II A theory \cite {witt} has been obtained. 

The procedure elaborated for one-fourth BPS states can be generalised
to construct supergravity solution for other non-planar 
networks \cite{ahar,sand}. The challenging problem of  finding
closed form expression for localised/partially localised solutions 
still remains.  
\vskip1cm

{\bf Acknowledgements:} I would like to thank Alok Kumar for
the fruitful discussions during the initial stages of this work.
I am grateful to Ashoke Sen for his comments and suggestions. 
I would like to thank Donald Marolf for explaining the 
difficulties in finding localised solutions. 
I would like to thank the organisers of the Black Hole
Conference, Trieste  where some of the issues on
localised and delocalised solutions got clarified.
I owe my thanks to
Department of Theoretical Physics, TIFR for providing
me local hospitality where part of this work was done.
I would like to thank CSIR for the grant.


\end{document}